\documentclass{mic2015}

\begin{document}

\title{On the Minimum Labelling Spanning \emph{bi}-Connected Subgraph problem}

\author{Jos\'{e} Andr\'{e}s Moreno P\'{e}rez\inst{1} \and Sergio Consoli\inst{2}}

\institute{
  Department of Computing Engineering, Universidad de La Laguna, Tenerife, Spain\\
  \email{jamoreno@ull.edu.es}
  \and
  ISTC/STLab, National Research Council (CNR), Catania, Italy\\
  \email{sergio.consoli@istc.cnr.it}
}

\id{id}

\maketitle

\begin{abstract}
We introduce the minimum labelling spanning bi-connected subgraph problem (MLSBP)
replacing connectivity by bi-connectivity in the well known
minimum labelling spanning tree problem (MLSTP).
A graph is bi-connected if, for every two vertices, there are, at least, two vertex-disjoint paths joining them.
The problem consists in finding the spanning bi-connected subgraph or block with minimum set of labels.
We adapt the exact method of the MLSTP to solve the MLSTB and the basic greedy constructive heuristic, the maximum vertex covering algorithm (MVCA).
This procedure is a basic component in the application of metaheuristics to solve the problem.
\end{abstract}

\section{Introduction}
\label{Introduction}

A labelled undirected graph is a graph where each edge has a colour or a label from a finite set of labels.
A well known problem in labelled graphs is the MLSTP consisting of finding the spanning tree that uses the minimum set of labels \cite{MLSTP-intell-ASOC}.
Other labelling problems has been proposed in the literature, some derived from the MLSTP.
One of the most recently studied is the $k$LSFP where the objective is to find forests, therefore relaxing the connectivity property \cite{VNSkLSF2015}.
We propose, in the opposite direction, to deal with bi-connectivity instead of connectivity.
An undirected graph is bi-connected if any pair of vertices are joined by two vertex-disjoint paths.
Equivalently, it is a graph that remains connected by dropping any single vertex.
The bi-connected graphs or networks have important applications in robustness of transport, communication and social networks.
This condition guarantees that such networks remain connected in the event of a node failure.
The shortest bi-connected network joining a set of vertices is a cycle or a ring passing throws all of them. For this reason the first computer networks in the 80's were based on rings.

\section{Bi-connectivity}

An undirected graph $G$ is connected if and only if for any pair of vertices there is, at least, a joining path.
The connected components of a graph are the maximal connected subgraphs.
The connected components of a graph constitute partitions of its set of of vertices and edges.
A connected graph has only one connected component. A spanning tree is obtained by iteratively removing edges from the connected graph 
until no 
cycles exist. 

In the other hand, an undirected graph $G$ is bi-connected if and only if for any pair of vertices there are, at least, two vertex-disjoint joining paths (i.e. two paths without a common vertex, excluding the joined terminal pairs).
A cut vertex or articulation point is a vertex whose removal disconnects the graph.
A graph is bi-connected if it has not cut vertex.
The bi-connected components or blocks of a graph $G$ are the maximal bi-connected subgraphs of $G$.
A cut vertex belongs to more than one block; therefore the blocks are not disjoints in terms of sets of vertices.
On the other side, the blocks are disjoint as set of edges, but an edge could not be in any block.
An edge that joints two different blocks is a bridge.
Every edge of the graph joins two vertices of the same block; otherwise it is a bridge.
The set of bridges and blocks (as set of edges) constitutes a partition of the set of edges.
An isolated vertex is assumed to be bi-connected.


\section{Formulation of the problem}

A labelled undirected graph $(G,\cal{L})$ = $(V,E,\cal{L})$ consists in an undirected graph $G = (V,E)$  and a finite set of labels $\cal{L}$ where every edge $e\in E$ has an unique label $l(e)\in \cal{L}$.
Several labelling problems have been defined in the literature. 
The well-known minimum labelling spanning tree (MLST) problem (Chang and Leu, 1997; Krumke and Wirth, 1998) consists of finding the spanning tree $T^*$ with minimum number of different labels.
Given a spanning tree $T = (V,F)$ of $G$, let $L(T) = \{ l \in \cal{L}$$| \exists e \in F, L(e) = l\})$.
The MLST is the spanning tree $T^*$ that minimizes $|L(T)|$.
The problem can be alternatively formulated terms of set of labels.
Given a subset of labels $L \subseteq \cal{L}$, the corresponding graph is $G(L)$ = $(V,E(L))$
where $E(L) = \{ e \in E | L(e)\in \cal{L} \}$.
The problem is to determine the set of labels such that the corresponding graph is connected and has the minimum possible number of labels.
That is the goal is to find the set $L^* \subseteq \cal{L}$ that minimizes $|L|$, where $L \subseteq \cal{L}$ and $G(L)$ is connected.
To get the spanning tree, iteratively edges are removed until no cycles exist.

We now introduce the minimum labelling spanning block (MLSB) problem, which consists of finding the spanning bi-connected subgraph $B^*$ having the minimum number of distinct labels.
Alternatively, the problem is to find the set of labels such that the corresponding graph is bi-connected with minimum number of labels.
It is to find the set $L^* \subseteq \cal{L}$ that minimizes $|L|$, where $L \subseteq \cal{L}$ and $G(L)$ is bi-connected.

\section{Solution algorithms}
\label{sec_algortithms}

The exact approach for the MLST problem \cite{MLSTP-EJOR} is adapted to solve the MLSB problem.
The method is based on an $A^*$ or backtracking procedure to test the subsets of $\cal{L}$.
The algorithm performs a branch and prune procedure in the partial solution space based on a recursive procedure that attempts to find a better solution from the current incomplete solution.
The main program of the exact method calls the recursive procedure with an empty set
of labels, and iteratively stores the best solution to date, say $L^*$.
The key procedure of the method is a subroutine that, for a given set of labels $L \subseteq \cal{L}$, determines if the graph $G(L)$ is connected or not.
To adapt the method to the MLSB problem, this procedure is replaced by a subroutine to determine if $G(L)$ is bi-connected or not. 
The number of sets tested, and therefore the running time, can be shortened by pruning the search tree using simple rules.
For example DFS (Depth First Search) can be used to determine 
bi-connectivity of the graph. 
Graph traversal algorithms \cite{Sedg2011} like DFS and BFS (Breath First Search) are strategies, linear in the number of edges, which visit the vertices of a graph by following the path leaded by its edges, backtracking as they encounter dead-ends. They are the basis for many graph-related algorithms, including topological sorts, planarity, and connectivity testing.
DFS uses the rule \emph{``first deep and then wide''}, meaning that it visits the child nodes before visiting the sibling nodes; that is, it traverses the depth of any particular path before exploring its breadth. 
DFS may be used to determine if a graph is connected and, in the positive case, to obtain its connected components.
At this purpose, DFS starts iteratively from a non visited vertex, until all the vertices are then visited.
The vertices (and edges) of each connected components are those visited in each run of the DFS.
The number of connected components is the number of runs of the DFS.
BFS instead uses the rule \emph{``first wide and then deep''}, meaning that 
it visits the parent nodes before visiting the child nodes, and a queue is used in the search process. BFS if often used to determine the shortest path from the root to the rest of the vertices.

DFS is extended in our exact method to determine if a graph is bi-connected and, in the positive case, to obtain its bi-connected components or blocks \cite{HT73}.
DFS identifies the cut vertex by considering the tree generated by the search process.
This tree rooted in the starting vertex consists in the edges traversed from a visited vertex to the next visited one; 
the rest of the edges are back edges.
The root of the DFS tree is a cut vertex if it has more than one outgoing tree edge.
A vertex $v$, which is not the root of the DFS tree, is a cut vertex if it has a child $w$ such that no
back edge starting in the subtree of $w$ reaches an ancestor of $v$.
It is determined during the execution of the DFS by using the order in which the vertices are visited.
The DFS algorithm recursively get the first visited node 
that is reachable from $v$ by using a directed path
having at most one back edge. 
If that vertex was not visited before $v$ by DFS, then $v$ is a cut vertex.
DFS visits the vertices and edges of each bi-connected component consecutively.
Therefore, the blocks are obtained using a stack to keep track of the block being traversed by DFS.

The MVCA (Maximum Vertex Covering Algorithm) is one of the first heuristic for MLSTP and is used as basic component of many metaheuristics applied to the problem.
It is a greedy constructive algorithm that selects the labels using the number of connected components as greedy function to minimize. 
The MVCA for the MLSBP starts with the empty set of labels, and then each vertex is evaluated as starting block and connected component.
Then the algorithm iteratively adds a label to the partial subgraph, selected by using the number of blocks plus the number of connected components 
as greedy function to minimize.
To know the number of blocks and components when a label is added, the corresponding edges are included one by one.
Note that each block is in only one connected component, so every component has its own block.
If connected components and blocks are known, when an edge is added the new number of blocks and components of the resulting graph are computed as follows.
First, if the edge joins two vertices of the same block, then the connected components and blocks do not change in number.
Second, if the edge joins two vertices of different connected components, then they are joined into a single connected component including the blocks of both components, where the added edge plays the role of a bridge for the new graph.
Finally, if the edge joins two vertices of different blocks into the same connected component, 
then the number of components does not change, but the blocks 
are joined into a single block belonging to this component. 
These are the blocks that are traversed by the shortest path joining the two extreme vertices of the included edge.
This path may be obtained by any BFS algorithm linearly with the number of edges. 

The partial solution at each iteration of the MVCA is constituted by the set of labels already included in the resulting subgraph. 
Then the label that most reduces the number of blocks and connected components at that step is selected to be included in the partial solution, and consequently the resulting subgraph is updated by adding the edges associated to the selected label.
Note that the inclusion of an edge may reduce the number of blocks (by different amounts), the number of connected components (exactly by one), or neither of the two cases; but never both.
The adding of a label instead may reduce the number of blocks, the number of connected components, both or neither of them, and in different amounts.

\section{Summary and Outlook}

We introduce the MLSBP and adapt to it the exact solution method and MVCA used for the MLSTP.
In the incoming future we plan to implement and compare other successful metaheuristics derived from the MLSTP literature. 
In order to consider real-world applications, we will extend the problem by considering that an edge can have associated more than one label, where each label may represent a different company in a transportation network perspective, or a different communication mode, frequency, or wavelength in telecommunication networks, and that each pair of nodes can be connected by one or more such multi-labeled connecting edges.

\begin{footnotesize}
\bibliography{bibliography}
\bibliographystyle{plain}
\end{footnotesize}

\end{document}